\documentclass[%
 reprint,
superscriptaddress,
 amsmath,amssymb,
 aps,
 prl
]{revtex4-2}

\usepackage{xcolor}
\usepackage{graphicx}
\usepackage{dcolumn}
\usepackage{bm}

\begin{document}


\title{\textbf{Pressure-Driven Moiré Potential Enhancement and  Tertiary Gap Opening\\ in Graphene/h-BN Heterostructure 
} 
}%

\author{Yupeng Wang}
\thanks{These authors contributed equally to this work.}
\affiliation{Department of Physics, University of Science and Technology of China, Hefei, Anhui 230026, China}
\affiliation{Deep Space Exploration Laboratory, University of Science and Technology of China, Hefei, Anhui 230026, China}

\author{Jiaqi An}
\thanks{These authors contributed equally to this work.}
\affiliation{Department of Physics, University of Science and Technology of China, Hefei, Anhui 230026, China}
\affiliation{International Center for Quantum Design of Functional Materials, University of Science and Technology of China, Hefei, Anhui 230026, China}
\affiliation{CAS Key Laboratory of Strongly Coupled Quantum Matter Physics, University of Science and Technology of China, Hefei, Anhui 230026, China}

\author{Chunhui Ye}
\thanks{These authors contributed equally to this work.}
\affiliation{Department of Physics, University of Science and Technology of China, Hefei, Anhui 230026, China}
\affiliation{Deep Space Exploration Laboratory, University of Science and Technology of China, Hefei, Anhui 230026, China}

\author{Xiangqi Wang}
\affiliation{Jihua Laboratory Testing Center, Jihua Laboratory, Foshan 528000, China}

\author{Di Mai}
\affiliation{Department of Physics, University of Science and Technology of China, Hefei, Anhui 230026, China}
\affiliation{Deep Space Exploration Laboratory, University of Science and Technology of China, Hefei, Anhui 230026, China}

\author{Hongze Zhao}
\affiliation{Department of Physics, University of Science and Technology of China, Hefei, Anhui 230026, China}
\affiliation{Deep Space Exploration Laboratory, University of Science and Technology of China, Hefei, Anhui 230026, China}

\author{Yang Zhang}
\affiliation{Department of Physics, University of Science and Technology of China, Hefei, Anhui 230026, China}
\affiliation{Deep Space Exploration Laboratory, University of Science and Technology of China, Hefei, Anhui 230026, China}

\author{Chiyu Peng}
\affiliation{Department of Physics, University of Science and Technology of China, Hefei, Anhui 230026, China}
\affiliation{Deep Space Exploration Laboratory, University of Science and Technology of China, Hefei, Anhui 230026, China}

\author{Kenji Watanabe}
\affiliation{National Institute for Materials Science, Tsukuba, Japan}

\author{Takashi Taniguchi}
\affiliation{National Institute for Materials Science, Tsukuba, Japan}

\author{Xiaoyu Sun}
\affiliation{The Centre for Physical Experiments, University of Science and Technology of China, Hefei, Anhui 230026, China}

\author{Rucheng Dai}
\affiliation{The Centre for Physical Experiments, University of Science and Technology of China, Hefei, Anhui 230026, China}

\author{Zhongping Wang}
\affiliation{The Centre for Physical Experiments, University of Science and Technology of China, Hefei, Anhui 230026, China}

\author{Wei Qin}
\email{Corresponding author: qinwei5@ustc.edu.cn}
\affiliation{International Center for Quantum Design of Functional Materials, University of Science and Technology of China, Hefei, Anhui 230026, China}
\affiliation{CAS Key Laboratory of Strongly Coupled Quantum Matter Physics, University of Science and Technology of China, Hefei, Anhui 230026, China}
\affiliation{Hefei National Laboratory, University of Science and Technology of China, Hefei 230088, China}

\author{Zhenhua Qiao}
\email{Corresponding author: qiao@ustc.edu.cn}
\affiliation{Department of Physics, University of Science and Technology of China, Hefei, Anhui 230026, China}
\affiliation{International Center for Quantum Design of Functional Materials, University of Science and Technology of China, Hefei, Anhui 230026, China}
\affiliation{CAS Key Laboratory of Strongly Coupled Quantum Matter Physics, University of Science and Technology of China, Hefei, Anhui 230026, China}
\affiliation{Hefei National Laboratory, University of Science and Technology of China, Hefei 230088, China}

\author{Zengming Zhang}
\email{Corresponding author: zzm@ustc.edu.cn}
\affiliation{Deep Space Exploration Laboratory, University of Science and Technology of China, Hefei, Anhui 230026, China}
\affiliation{CAS Key Laboratory of Strongly Coupled Quantum Matter Physics, University of Science and Technology of China, Hefei, Anhui 230026, China}
\affiliation{The Centre for Physical Experiments, University of Science and Technology of China, Hefei, Anhui 230026, China}


\begin{abstract}
Moiré superlattices enable engineering of correlated quantum states through tunable periodic potentials, where twist angle controls periodicity but dynamic potential strength modulation remains challenging. Here, we develop a high-pressure quantum transport technique for van der Waals heterostructures, achieving the ultimate pressure limit ($\sim 9~\text{GPa}$) in encapsulated moiré devices. In aligned graphene/h-BN, we demonstrate that pressure induces a substantial enhancement of the moiré potential strength, evidenced by the suppression of the first valence bandwidth and the near-doubling of the primary bandgap. Moreover, we report the first observation of a tertiary gap emerging above $6.4$ GPa, verifying theoretical predictions.  Our results establish hydrostatic pressure as a universal parameter to reshape moiré band structures. By enabling quantum transport studies at previously inaccessible pressure regimes, this work expands the accessible parameter space for exploring correlated phases in moiré systems.

\end{abstract}

\maketitle

Moiré superlattices, formed by twisted or lattice-mismatched van der Waals layers, exhibit extraordinary quantum phenomena through their tunable periodic potentials \cite{bn1,bn2,bn3,bn4}. The formation of ultra-flat electronic bands in these artificial crystals enables the emergence of exotic quantum states, including superconductivity \cite{bn5,bn6,bn7,bn8,bn9,bn10,bn11}, correlated insulators \cite{bn12,bn13,bn14,bn15,bn16,bn17,bn18,bn19,bn20,bn21}, and integer and fractional quantum anomalous Hall effects  \cite{bn22,bn23,bn24,bn25,bn26,bn27,bn28}. The structure of these moiré flat bands is governed by the interplay between the periodicity  and strength of the moiré potential \cite{bn3,bn4}. The effects of moiré periodicity have been extensively studied by varying twist angles, revealing that intriguing correlated phenomena are typically confined to a narrow range of twist angles. However, the fixed nature of twist angles after device assembly imposes significant challenges for cross-device comparisons due to inter-sample disorder arising from fabrication variations.

Hydrostatic pressure as a promising resolution  can continuously regulate interlayer coupling without altering moiré periodicity \cite{bn29,bn30,bn31,bn32}. Beyond modifying the moiré band structures in a manner analogous to tuning the twist angle \cite{bn3,bn33,bn34,bn35,bn36}, pressure mitigates the impact of inter-sample disorder by enabling continuous band structure evolution within a single device through \textit{in situ} pressure loading. Theoretical proposals further predict pressure-driven band topology reconstruction and enhanced electronic correlations \cite{bn33,bn34,bn35,bn36,bn37,bn38,bn39}. However, quantum transport studies of moiré devices have been restricted to $\leq 3$~GPa using conventional piston cells \cite{bn29,bn31,bn32}, far below the $\sim 9$~GPa threshold where h-BN encapsulation layers (essential for device integration)  undergo structural phase transitions \cite{bn29,bn30,bn41,bn43}.

Here we overcome this limitation by integrating diamond anvil cell (DAC) techniques with quantum transport measurements in h-BN-encapsulated moiré devices. Using the graphene/h-BN superlattice as a prototypical platform, we demonstrate a significant enhancement of the moiré potential and observe the emergence of a tertiary bandgap absent at ambient conditions. Our results open a new parameter space for engineering and exploring correlated and topological states in moiré systems.

\begin{figure*}[htbp]
  \centering
  \includegraphics[width=1 \textwidth]{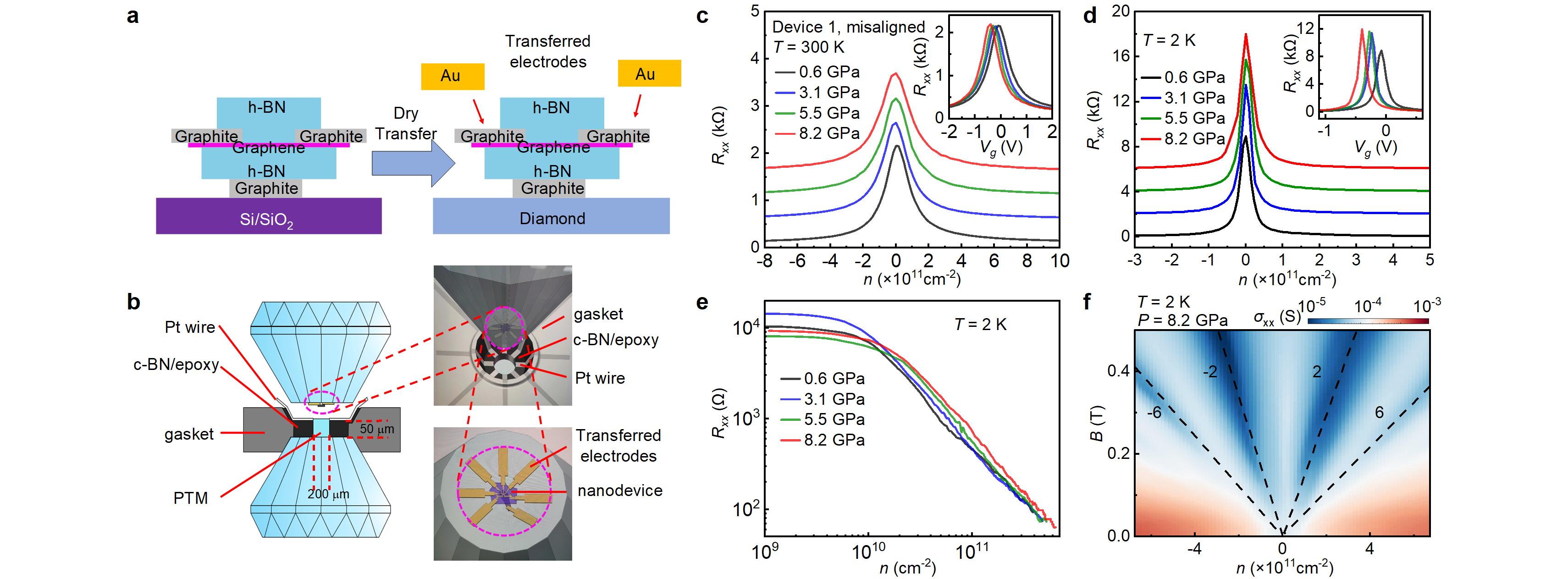} 
\caption{
(a) Left: Prefabricated device on a Si/SiO\textsubscript{2} substrate. 
Right: Transfer the device onto the diamond anvil surface. The gold electrodes are subsequently placed onto the graphite through the transferred electrode method. 
(b) Left: The DAC setup. The sample is mounted on the upper surface of the diamond, with platinum electrodes positioned on a tungsten gasket that has been pre-coated with a composite of cubic boron nitride and epoxy.  
Right:  Connecting Pt wires and the nanodevice via transferred electrodes. Eight Pt electrodes are present on the gasket, seven of which are used for this sample.
(c) Resistance $R_{xx}$ versus carrier density $n$ for monolayer graphene measured at $300~\text{K}$ under different pressures (offset:  0.5~k$\Omega$).
Inset: raw data of $R_{xx}$ versus gate voltage $V_g$. The $n-V_g$ relation is extracted from Quantum Hall measurements.
(d) $R_{xx}$ versus $n$ for monolayer graphene measured at $2~\text{K}$ under different pressures (offset:  2~k$\Omega$). Inset: raw data of $R_{xx}$ versus $V_g$. 
(e) Detailed Log-log plot of $R_{xx}$ versus $n$ at $2~\text{K}$. 
(f) Landau fan diagram of monolayer graphene at $2~\text{K}$ and $8.2~\text{GPa}$. 
}
  \label{fig:1}
\end{figure*}

To address the challenges of high-pressure quantum transport measurements in moiré devices, we developed two key innovations, as shown in Fig. 1(a). First, a \lq transfer after fabrication\rq~ technique enables the integration of h-BN-encapsulated graphene devices on diamond anvils: devices pre-fabricated on silicon via standard lithography are dry-transferred onto diamond surfaces \cite{bn30,bn41,bn42}, preserving precise sample geometry and uniform quality. Second, a \lq transferred electrode\rq ~method allows for direct deposition of pre-formed gold electrodes to specific device regions, bypassing the need for conventional lithography on diamond substrates. This approach ensures robust electrical contacts while maintaining sample integrity under extreme pressures. 

Figure~1(b) illustrates the integrated pressure cell design. The nano-device, combined with a graphite bottom gate, is transferred onto the upper diamond anvil (type IIA, 500~$\mu$m culet). Subsequently, Au electrodes are transferred to specific positions to electrically connect with the nano-device. Pt wires (thickness~$\sim$5~$\mu$m) are placed on the gasket of the lower diamond surface. The gasket, pre-coated with c-BN/epoxy composite for insulation, incorporates a laser-drilled pressure chamber with a diameter of 200~$\mu$m. The chamber is filled with oil as the pressure-transmitting medium to ensure uniform pressurization of the sample. Upon pressurization, the transferred Au electrodes on the upper diamond are tightly pressed against the Pt wires located on the gasket, ensuring robust electrical contact. The eight Pt wires positioned on the gasket provide eight independent measurement channels. This design enables \textit{in situ} transport measurements under hydrostatic pressures up to the phase transition threshold of device encapsulation layers, while operating at temperatures below 2~K and in magnetic fields exceeding 9~T.

To evaluate the reliability of our technique, we perform benchmark transport measurements on a monolayer graphene device (Device 1), where the graphene and h-BN encapsulation layers are intentionally misaligned to avoid formation of moiré superlattice. Figure~1(c) shows the room-temperature resistance as a function of the carrier density for the device under different pressures. At $0.6~\text{GPa}$, the resistance behavior closely aligns with that of previously reported ambient devices \cite{bn44,bn45,bn46,bn47,bn48}. Notably, increasing the pressure up to 8 GPa has a minimal impact on the resistance for a given carrier density. At 2 K, the resistance peak at the Dirac point exhibited a full width at half maximum of $3.0 \times 10^{10} \, \mathrm{cm}^{-2}$ (Fig. 1(d)), indicating low charge disorder across all pressures studied. A double-logarithmic plot of resistance versus carrier density shows a residual carrier density of less than $10^{10} \, \mathrm{cm}^{-2}$, which remains unaffected by pressure variations (Fig. 1(e)). Figure 1(f) presents a Landau fan diagram at $8.2~\text{GPa}$ and $2~\text{K}$, where clear evidence of integer quantum Hall states is observed below $0.2~\text{T}$, further  confirming that the device is of high quality. These results confirm that the monolayer graphene device retains its exceptional quality under pressures up to $8.2~\text{GPa}$, establishing the reliability of our methodology in investigating physical properties that rely on the high sample quality of devices (See Supplemental Material \cite{sm} for raw data and application to other moiré systems like twisted bilayer graphene (TBG)).

\nocite{sm1, sm3, sm4, sm5, sm6,sm8,sm9,sm10,sm11,sm12,sm13}

We choose the aligned graphene/h-BN moir\'{e} superlattice as a prototypical system to demonstrate the pressure-induced modulation of the moir\'{e} potential strength. The devices are fabricated by aligning the top h-BN layer with monolayer graphene while keeping the bottom h-BN layer misaligned. Figures~2(a) and 2(b) present the longitudinal resistance $ R_{xx} $ as a function of carrier density $ n $ for Device~2 and Device~3, respectively, under varying pressures. The resistance peak at $ n = 0 $, originating from the primary Dirac point (PDP), persists across all pressure conditions. At the first pressure, Device~2 (Device~3) exhibits resistance peaks at carrier densities of $ \pm 2.8 \times 10^{12}~\text{cm}^{-2} $ ($ \pm 2.5 \times 10^{12}~\text{cm}^{-2} $), arising from superlattice-induced secondary Dirac points (SDPs) and corresponding to a twist angle of $ 0.4^\circ $ ($ 0.1^\circ $) between graphene and the top h-BN. Under applied pressure, the carrier density at these SDPs remains fixed while the gate voltage $ V_g $ at the SDPs decreases with pressure, indicating that pressure enhances gate capacitance without altering the twist angle (see Supplemental Material \cite{sm} for detailed capacitive analysis).

\begin{figure}[htbp]
  \centering
  \includegraphics[width=0.5 \textwidth]{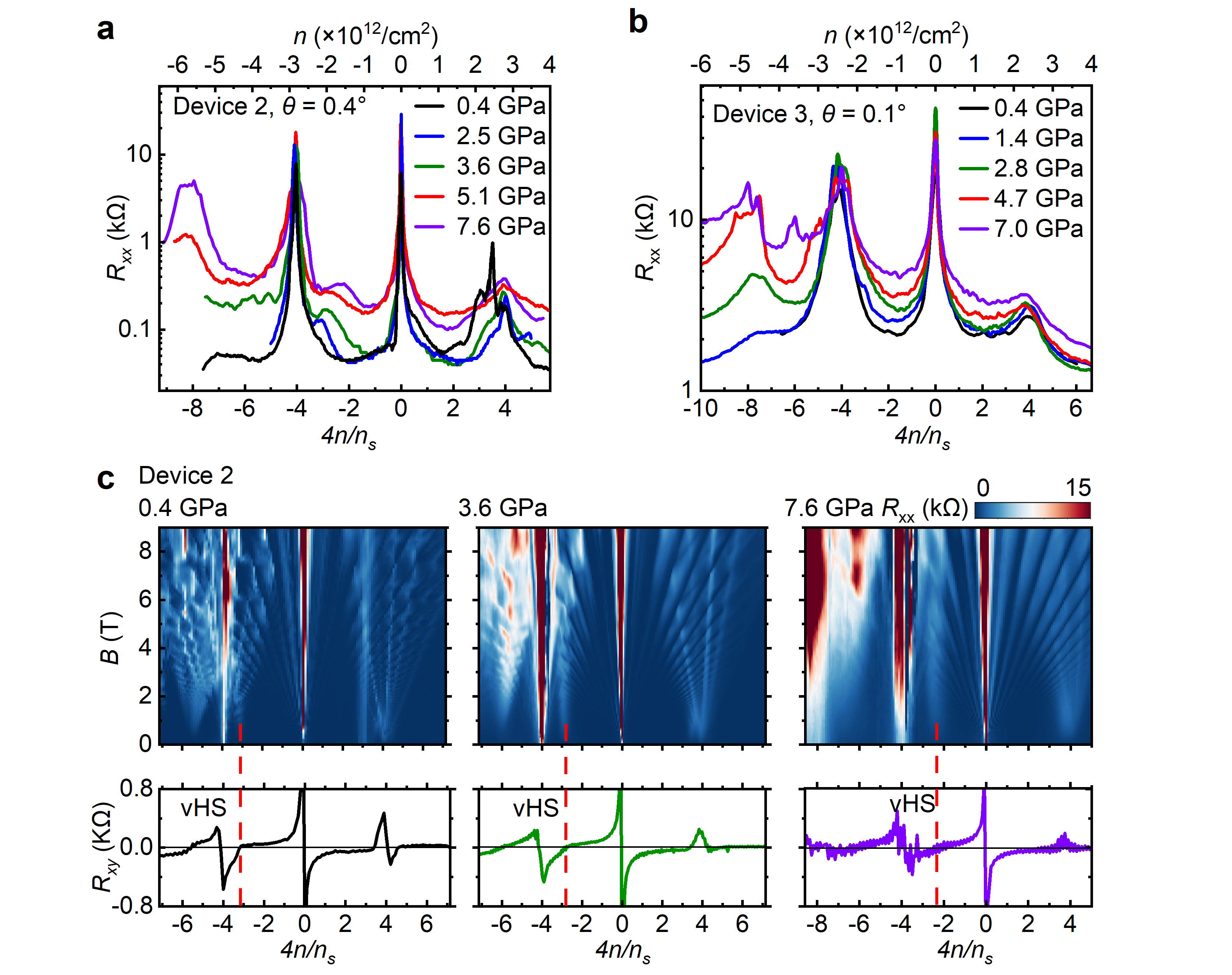} 
\caption{Longitudinal resistance $R_{xx}$ of the graphene/h-BN moir\'{e} superlattice as a function of the carrier density $n$ and moir\'{e} filling factor $n/n_{s}$ ($n_s$ corresponds to full filling of the first moir\'{e} band) at different pressures for: (a) device 2 and (b) device 3 . 
(c) Top: Landau fan diagrams measured for device 2 at pressures of $0.4~\text{GPa}$ (left), $3.6~\text{GPa}$ (middle), and $7.6~\text{GPa}$ (right). 
Bottom: Transverse resistance $R_{xy}$ versus $n$ at a magnetic field of 0.5~T for pressures of $0.4~\text{GPa}$ (left), $3.6~\text{GPa}$ (middle), and $7.6~\text{GPa}$ (right). 
 The vHSs are indicated by the red dashed lines. 
All the data are obtained at $2~\text{K}$.}

  \label{fig:2}
\end{figure}

Next, we focus on data from Device 2, with similar behaviors observed in Device 3. As shown in Fig. 2(a), the resistance for a given carrier density within the metallic regime increases with increasing pressure, suggesting a continuous reduction of the moiré bandwidth. Specifically, resistance is given by $R \sim m^*/n e \tau$ within the Drude model, where $m^*$ denotes the effective mass of the charge carrier, $n$ is the carrier density, and $\tau$ is the relaxation time. For a given device, it is reasonable to assume that $\tau$ remains unchanged at different pressures, as contributions to $\tau$ are dominated primarily by disordered scattering at low temperatures. Consequently, the increase in resistance reflects mainly an increase in $m^*$, which is typically inversely proportional to the bandwidth. Furthermore, the resistance kinks observed between PDP and SDP are attributed to the van Hove singularity (vHS) of the first moiré valence band, as evidenced by the sign change of the Hall signal extracted at $0.5\,\mathrm{T}$ and depicted in Fig. 2(c). The carrier densities at the vHSs are $-2.2 \times 10^{12}\,\mathrm{cm}^{-2}$ and $-1.6 \times 10^{12}\,\mathrm{cm}^{-2}$ for $0.4\,\mathrm{GPa}$ and $7.6\,\mathrm{GPa}$, respectively. With increasing pressure, the vHS shifts towards the charge neutrality point, indicating that the band dispersion can be modulated by pressure.

To further demonstrate the tailoring of moiré potential strength under pressure, we extract the bandgaps by measuring the temperature dependence of the conductivity at different pressures. Figures 3(a) and 3(b) show the Arrhenius plots for conductivity at the primary and secondary Dirac points, where primary gap ($\Delta_\mathrm{P}$) and  secondary gap  ($\Delta_\mathrm{S}$) are obtained through linear fittings within the thermally activated regimes (solid lines). As depicted in Fig. 3(d), $\Delta_\mathrm{P}$ exhibits nearly linear behavior with increasing pressure, increasing from approximately $25(4)\,\mathrm{meV}$ at $0.4\,\mathrm{GPa}$ to around $49(3)\,\mathrm{meV}$ at $5.1\,\mathrm{GPa}$. Moreover, $\Delta_\mathrm{S}$ remains nearly unchanged, with a value of $\sim \! 20\,\mathrm{meV}$ from $0$ to $2.5\,\mathrm{GPa}$, which is consistent with previous reports \cite{bn29}. By further increasing the pressure, we find that $\Delta_\mathrm{S}$ starts to rise from $\sim 2.5~\mathrm{GPa}$.

\begin{figure}[htbp]
  \centering
  \includegraphics[width=0.5 \textwidth]{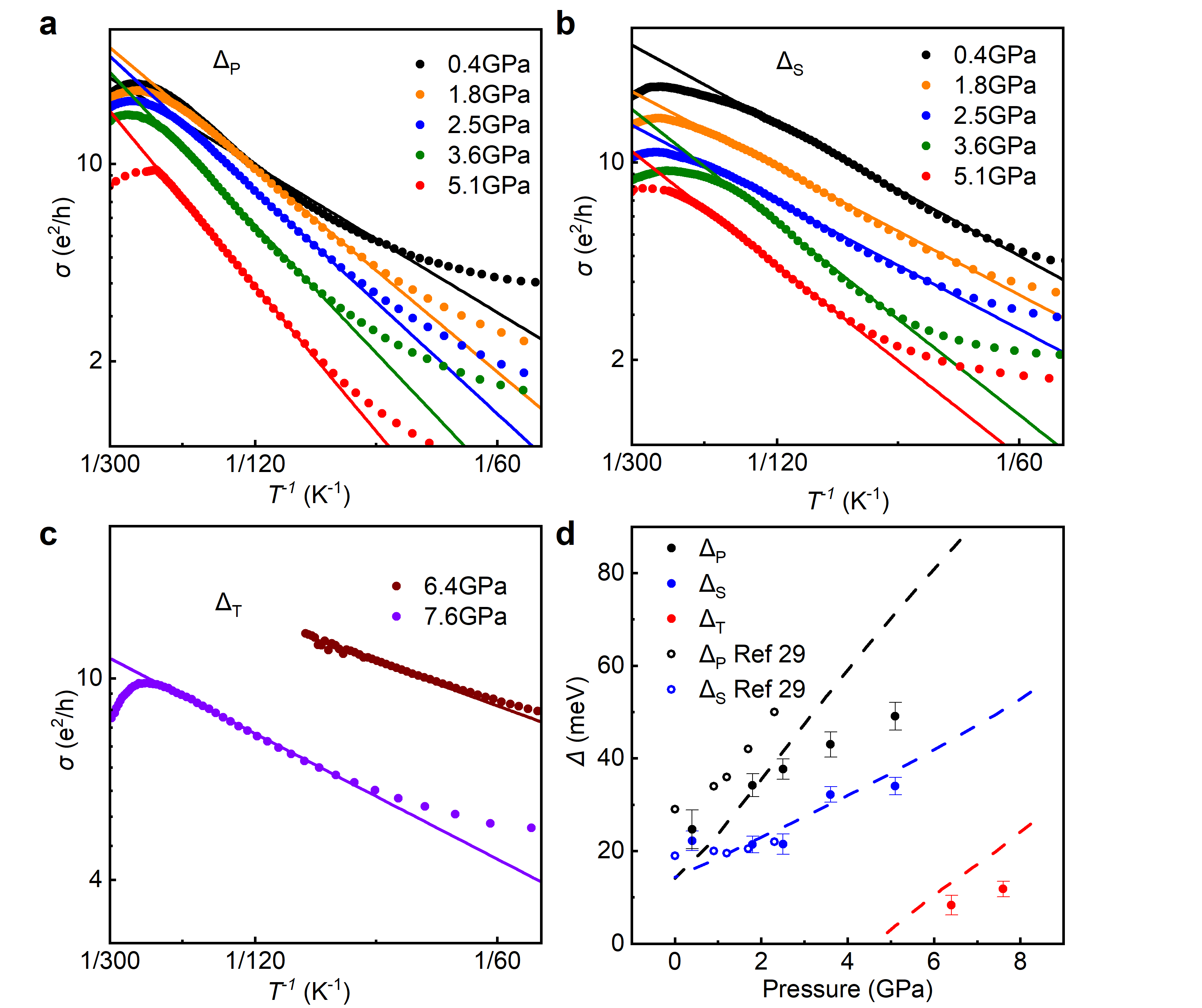} 
\caption{Arrhenius plots of the conductivity $\sigma$ as a function of the inverse temperature $1/T$ for the primary gap (a), secondary gap (b), and tertiary gap (c) at different pressures.  The lines are corresponding fittings to these data.
(d) Bandgaps as a function of pressure, with the dashed curves representing predictions from the fully relaxed tight-binding model. 
The open circles in (d) are the results from Ref. \cite{bn29}. }
  \label{fig:3}
\end{figure}

The primary gap opening is attributed to the breaking of inversion symmetry resulting from a nonzero average mass term derived from moiré strains, as evidenced by experimental observations \cite{bn29,bn53,bn54,bn55}. It was reported that $\Delta_\mathrm{P}$ is proportional to both interlayer coupling strength and strain magnitude \cite{bn56}. Consequently, as the pressure increases, the interlayer coupling between graphene and h-BN, as well as the associated strain in graphene, enhances, leading to a rapid increase of $\Delta_\mathrm{P}$. In contrast, the secondary gap $\Delta_\mathrm{S}$ originates from the intricate interplay between interlayer coupling and strain \cite{bn29}, which can explain its unchanged behavior at lower pressures. The observed enhancement of $\Delta_\mathrm{S}$ at high pressures indicates that $\Delta_\mathrm{S}$ is predominantly determined by the increased interlayer coupling strength. The enhancements of both $\Delta_\mathrm{P}$ and $\Delta_\mathrm{S}$ strongly suggest that the pressure significantly increases the strength of the moiré potential.

In addition to the resistive peaks observed at PDP and SDP, as illustrated in Fig. 2(b), a third resistive peak emerges at a hole carrier density of approximately $-5.6 \times 10^{12} \, \mathrm{cm}^{-2}$ when the applied pressure exceeds $5.1\,\mathrm{GPa}$. This peak may be associated with a gap opening at the tertiary Dirac point (TDP), a phenomenon not yet observed at ambient pressure, even in perfectly aligned graphene/h-BN devices \cite{bn57}. As shown in Fig. 2(c), the Landau fan diagram measured at $7.6\,\mathrm{GPa}$ reveals high-resistivity states at TDP under a strong magnetic field, further confirming the presence of a global band gap $\Delta_T$. Moreover, as depicted in Fig. 3(c), Arrhenius analysis of the temperature dependence of the conductivity indicates that the band gap $\Delta_\mathrm{T}$ at TDP is approximately $8(2)\,\mathrm{meV}$ and $12(2)\,\mathrm{meV}$ for pressures of $6.4\,\mathrm{GPa}$ and $7.6\,\mathrm{GPa}$, respectively.

It was demonstrated that both out-of-plane corrugation and in-plane strain play crucial roles in the formation of bandgaps of graphene/h-BN moiré superlattice \cite{bn29,bn56}. To further understand the modulation of electronic properties of graphene/h-BN under high pressures, we first employ a large-scale atomic/molecular massively parallel simulator to simulate the lattice relaxation dynamics (See Supplemental Material \cite{sm} for details of the theoretical model). As shown in Fig. 4(a),  pressure reduces the interlayer spacing across all stacking regions. To minimize the stacking potential on the h-BN substrate, carbon atoms relax both in-plane and out-of-plane.  Different stacking regions exhibit distinct responses to pressure, resulting in out-of-plane corrugation and in-plane strain. With increasing pressure, these relaxation effects are enhanced, as illustrated in Figs.~4(a) and 4(b), profoundly modifying the moiré band structure.

\begin{figure*}[htbp]
  \centering
  \includegraphics[width=1 \textwidth]{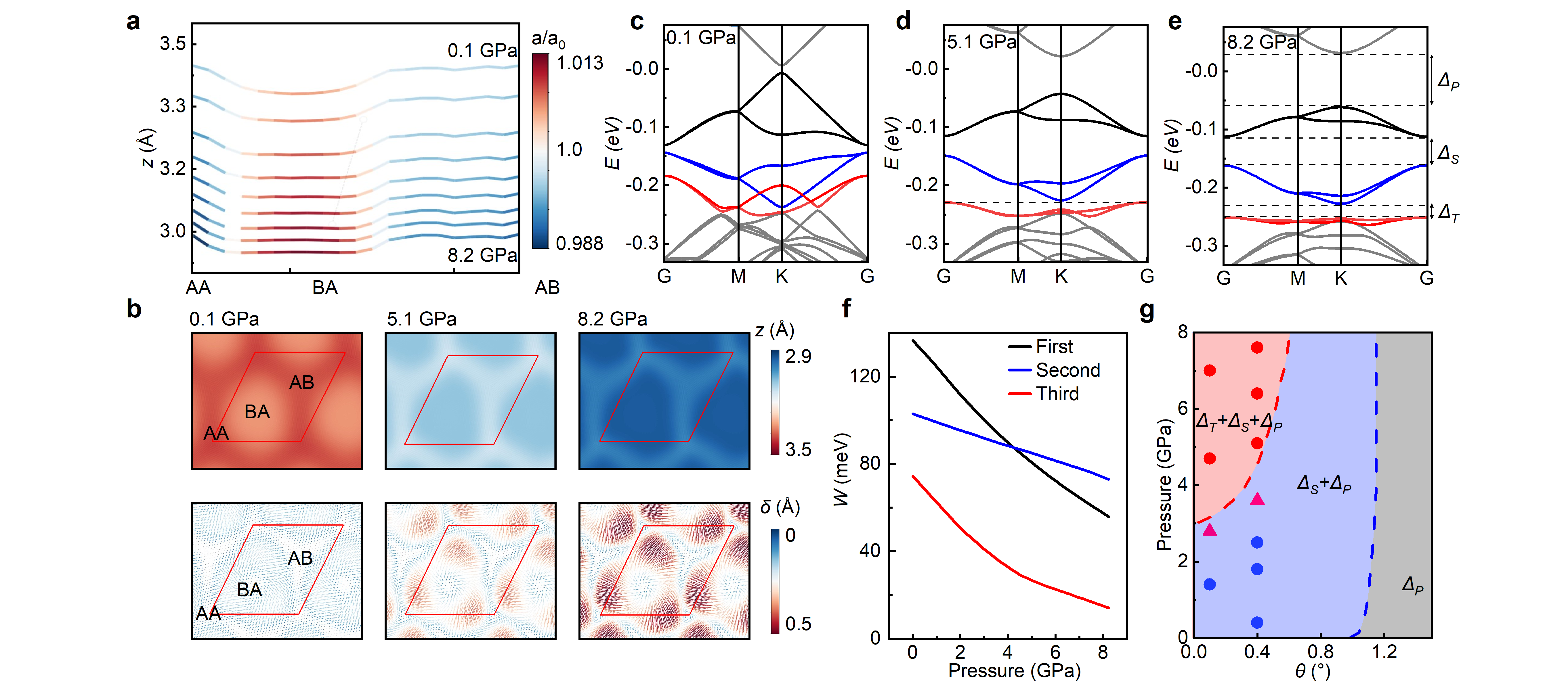} 
\caption{
(a) Pressure dependence of the interlayer distance, where the color bar denotes the magnitude of the in-plane strain.
(b) Interlayer distance $z$ (top) and in-plane relaxation $\delta$ (bottom) at different pressures. Calculated band structures at 0.1~GPa (c), 5.1~GPa (d) and 8.2~GPa (e), respectively. 
(f) Pressure dependence of the band width $W$ for the first, second, and third mori\'{e} valence bands. (g) Phase diagram with graphene/h-BN with varying $\theta$ from $0^\circ$ to $1.5^\circ$ and pressures up to $8~\text{GPa}$. Experimental data points with a sizeable $\Delta_T$, signature of $\Delta_T$, and no $\Delta_T$ represented by red, pink, and blue colors, respectively. }
  \label{fig:4}
\end{figure*}

Next, we input the fully relaxed structures into a tight-binding model to calculate the band structures at different pressures. As summarized by the dashed lines in Fig. 3(d), the calculated bandgaps at PDP and SDP are comparable with the experimental findings. Both gaps increase monotonically with increasing pressure. Notably, as shown in Fig. 4(f), the bandwidth of the first valence band rapidly decreases as the pressure increases, decreasing from approximately $140$ meV at $0.0$ GPa to $50$ meV at $8.2$ GPa. As shown in Fig. 4(d), our band structure calculation at $5.1$ GPa reveals an indirect band gap of $\sim~3$ meV between the second and third moiré valence bands. In contrast to the bandgaps at PDP and SDP, the emergence of the tertiary gap arises from the pressure-induced reduction of the bandwidths, lifting the overlap between the second and third moiré valence bands when the pressure exceeds $\sim~5.0$ GPa, as illustrated in Figs. 4(c) and 4(d). Moreover, as shown in Fig. 3(d), the theoretically obtained tertiary gap enlarges with the increase of pressure, with sizes being consistent with the experimental results.

We further explore the dependence of $\Delta_T$ on both twist angle and pressure. $\Delta_T$ emerges within a finite range of twist angles ($0^{\circ} \leq \theta \leq 0.7^{\circ}$), where the critical pressure ($P_{\mathrm{c}}$) required to open $\Delta_T$ increases monotonically with $\theta$ (Fig.~4(g)). Specifically, at $\theta = 0^{\circ}$, $\Delta_T$ appears at $P_{\mathrm{c}} \approx 3~\text{GPa}$, while for $\theta = 0.7^{\circ}$, the threshold rises to $P_{\mathrm{c}} \approx 9~\text{GPa}$. Beyond $\theta = 0.7^{\circ}$, no $\Delta_T$ emerges even under the maximum available pressure. This is because larger twist angles are generally associated with broader moiré bandwidths, thereby requiring higher pressure to suppress band overlap between the second and third moiré valence bands. This trend is confirmed by measurements on Device 3 with $\theta = 0.1^{\circ}$ as shown in Fig. 2(b): The resistivity peak associated with $\Delta_T$ emerges at a lower pressure of $2.8~\text{GPa}$. This peak is further strengthened at $4.7~\text{GPa}$, suggesting a sizable bandgap opening. In addition to the band gap, our calculation unveils a previously uncharted parameter space where $\Delta_T$ coexists with ultraflat bands (third valence bandwidth $\sim15~\text{meV}$ at $8.2~\text{GPa}$), offering a promising platform for exploring strong correlation effects within this band \cite{bn53,bn57,bn58,bn59,bn60}. Interestingly, we also observe an additional resistive feature emerging near the filling factor $\nu = -6$ under high pressure in both devices (Fig.~2b at 7.0~GPa and the Landau fan Fig.~2c at 7.6~GPa under high field). This feature occurs at half-filling of the second moiré valence band and potentially indicates a correlation-induced insulating state that warrants further investigation.

The high-pressure quantum transport platform developed here transcends the graphene/h-BN system. For example, in TBG, pressure facilitates the realization of flat bands at twist angles larger than the magic angle, increasing the density of states and Coulomb interaction strength \cite{bn33,bn34,bn35,bn61,bn62,bn63,bn64,bn65,bn66}. Simultaneously, pressure modulates the ratio of interlayer tunneling strengths between AA and AB stacking regions by altering atomic relaxation \cite{bn31,bn35}, a parameter frozen under ambient pressure but decisive for correlated states emergence \cite{bn67}. In twisted transition metal dichalcogenides, pressure-driven atomic corrugation and layer deformation can provide dynamic control over Berry curvature distribution and band topology, key ingredients for engineering isolated Chern bands and realizing fractional quantum anomalous Hall effect \cite{bn37,bn68}. Furthermore, in more general van der Waals heterostructures beyond those discussed above, our \textit{in situ} transport platform  can significantly strengthen proximity effects and potentially reveal novel interfacial electronic phenomena.

In summary, this work establishes hydrostatic pressure as a powerful tuning parameter to expand the accessible parameter space in the quantum transport study of moiré superlattices, broadening exploration of exotic states and potentially yielding a deeper understanding of the mechanisms underlying these intriguing correlated phenomena.

\begin{acknowledgments}
This work is financially supported by the National Key R\&D Program of China (Grant No. 2024YFA1408103), the National Natural Science
Foundation of China (No. 12074360, No. 12474158,
No. 12488101, and No. 12474134), the Anhui Initiative
in Quantum Information Technologies (AHY170000), the
Innovation Program for Quantum Science and Technology
(2021ZD0302800), and the Frontier Scientific Research
Program of Deep Space Exploration Laboratory under
Grant (No. 2022-QYKYJHHXYF-019) Device fabrication was partially carried out at the USTC Center for Micro- and Nanoscale Research and Fabrication. Measurements under high magnetic fields were supported by the Synergic Extreme Condition User Facility (SECUF). We also acknowledge the Supercomputing Center of USTC for providing high-performance computing resources.
\end{acknowledgments}


\bibliography{bngr}

\end{document}